\documentclass[12pt]{article}

\begin{document}

\begin{center}
{\Large\bf{}Relativistic tidal heating of Hamiltonian quasi-local
boundary expressions}
\end{center}

\begin{center}
Lau Loi So 
\end{center}

\begin{abstract}
Purdue and Favata calculate the tidal heating used certain
classical pseudotensors.  Booth and Creighton employed the
quasi-local mass formalism of Brown and York to demonstrate the
same subject. All of them give the result matched with the
Newtonian theory.  Here we present another Hamiltonian quasi-local
boundary  expressions and all give the same desired value. This
indicates that the tidal heating is unique as Thorne predicted.
Moreover, we discovered that the pseudo-tensor method and
quasi-local method are fundamentally different.
\end{abstract}

\section{Introduction}
In gravitation, one of the interested topics is that calculate the
tidal heating: the interaction between a nearly isolated
gravitating body and the external universe. Solar system provides
a typical example for the illustration: Jupiter and its satellite
Io~\cite{Peale}. Purdue~\cite{Purdue} and Favata~\cite{Favata}
examined the tidal heating for the classical pseudo-tensors. Booth
and Creighton used the quasi-local mass formalism of Brown and
York to demonstrate the same subject~\cite{Booth}. All of them
give the same value as the Newtonian perspective. Here we present
another Hamiltonian quasi-local boundary expressions to examine
the tidal heating, we find that the result is unique as Purdue
achieved, i.e., quasi-local expressions independent.

Thorne claimed that all pseudo-tensors give the same tidal work as
the Newtonian gravity~\cite{Thorne}. Nester realized that
pseudo-tensor method and quasi-local formalism are basically the
same~\cite{Nester1999}, however, Booth and Creighton prefer the
quasi-local method such that all quantities can be manipulated in
terms of real tensors on the quasi-local surface~\cite{Booth}.
Thus one may imagine that there is no surprise using quasi-local
expressions to give the same desired tidal work. Although using
the pseudo-tensor method and quasi-local method give the same
tidal heating rate, we find that the fundamental principle between
these two are different. In particular, the M$\o$ller
pseudo-tensor give the standard tidal heating~\cite{Favata} but
failed for inside matter requirement~\cite{CQGSoNesterChen2009}.
Meanwhile the Hamiltonian quasi-local method can rehabilitate this
handicap.

Generally speaking, different energy-momentum pseudo-tensor refers
to different gauge condition for the gravitational energy
localization, while different quasi-local boundary expression
focus on different boundary condition. Confined to the tidal
heating, we claim that the gauge condition and boundary condition
are elementarily the same terminology. Different gauge condition
corresponds different $E_{\rm{}int}$, where $E_{\rm{}int}$ is the
energy interaction between the isolated planet's quadrupolar
deformation and the external tidal field. We find that the tidal
heating remains unchange for different quasi-local boundary
expressions, thus the tidal heating is gauge invariant. However
there is a change: the exchangeable energy rate
$\dot{E}_{\rm{}int}$, they are gauge dependent. Here we explain
some terminology used in present paper. The expected tidal heating
or tidal work rate $\dot{W}=-\frac{1}{2}\dot{I}_{ij}E^{ij}$, where
$W$ refers to tidal work, the dot means differentiate w.r.t. time
t, $I_{ij}$ is the mass quadrupole moment of the isolated planet
and $E_{ij}$ is the tidal field of the external universe. Both
$I_{ij}$ and $E_{ij}$ are time dependent, but symmetric and trace
free. Here we emphasize that the tidal heating at which the
external field does work on the isolated body and this is an
energy dissipation process which means energy unexchangeable.
Conversely, there is an exchangeable process
$\dot{E}_{\rm{}int}\sim\frac{d}{dt}(I_{ij}E^{ij})$. Purdue uses
$E_{\rm{}int}=\frac{\beta+2}{10}I_{ij}E^{ij}$ to distinguish
different options how to localize the gravitational energy by
tuning the coefficient $\beta$~\cite{Purdue}.

\section{Technical background}
We used the same spacetime signature and notation as
in~\cite{MTW}: let the geometrical units $G=c=1$, where $G$ and
$c$ are the Newtonian constant and speed of light. The Greek
letters refer to the spacetime and Latin letters indicate the
spatial. For the idea of energy-momentum pseudo-tensor
$t_{\alpha}{}^{\mu}$, choose an appropriate super-potential
$U_{\alpha}{}^{[\mu\nu]}$
\begin{eqnarray}
\partial_{\nu}U_{\alpha}{}^{[\mu\nu]}=2\kappa\sqrt{-g}(T_{\alpha}{}^{\mu}+t_{\alpha}{}^{\mu}),\label{7aMay2015}
\end{eqnarray}
where $T_{\alpha}{}^{\mu}$ is the stress tensor,
$\partial_{\nu}U_{\alpha}{}^{[\mu\nu]}$ can be described as the
total energy-momentum complex and it is conserved since
$\partial^{2}_{\mu\nu}U_{\alpha}{}^{[\mu\nu]}\equiv0$. Recall the
Einstein equation: $G_{\mu\nu}=\kappa{}T_{\mu\nu}$, where
$G_{\mu\nu}$ is the Einstein tensor and $\kappa=8\pi{}G/c^{4}$.
Look closer at (\ref{7aMay2015}),
$\partial_{\nu}U_{\alpha}{}^{[\mu\nu]}$ consists two parts:
$2G_{\alpha}{}^{\mu}$ is the mass energy inside matter and
$t_{\alpha}{}^{\mu}$ is the gravitational energy-momentum in
vacuum. The component $t_{0}{}^{j}$ is the gravitational energy
flux density. The criterion of the interior mass-energy is
important. In particular, the classical M$\o$ller pseudo-tensor
cannot satisfy this inside matter
condition~\cite{CQGSoNesterChen2009}. Thus one can conclude that
M$\o$ller pseudo-tensor is not appropriate to describe the
energy-momentum in vacuum.

  The
gravitational tidal heating rate can be computed as
\begin{eqnarray}
2\kappa\dot{W}=\oint_{\partial{}V}\sqrt{-g}t_{0}{}^{j}\hat{n}_{j}\,r^{2}\,d\Omega,
\end{eqnarray}
where $r\equiv\sqrt{\delta_{ab}x^{a}x^{b}}$ is the distance from
the body in its local asymptotic rest frame and
$\hat{n}_{j}\equiv{}x_{j}/r$ is the unit radial vector.  In our
calculation, the metric tensor can be decomposed as
$g_{\alpha\beta}=\eta_{\alpha\beta}+h_{\alpha\beta}$ and its
inverse $g^{\alpha\beta}=\eta^{\alpha\beta}-h^{\alpha\beta}$. We
have the following physical expressions~\cite{Purdue}:
\begin{eqnarray}
h_{00}=\frac{2M}{r}+\frac{3}{r^{5}}I_{ab}x^{a}x^{b}-E_{ab}x^{a}x^{b},\quad{}
h_{0j}=\frac{2}{r^{3}}\dot{I}_{ij}x^{i}-\frac{10}{21}\dot{E}_{ab}x^{a}x^{b}x_{j}+\frac{4}{21}\dot{E}_{ij}x^{i}r^{2},
\label{27cMar2015}
\end{eqnarray}
note that $h_{ij}=\delta_{ij}h_{00}$.

\section{Hamiltonian quasi-local boundary expressions}
Geometric theories are invariant under local diffeomorphism.  Here
we review the Hamiltonian quasi-local boundary expressions from a
first order Lagrangian~\cite{SoIJMPD}:
\begin{equation}
{\cal{}L}=dq\wedge{}p-\Lambda(q, p),
\end{equation}
where $q$ and $p$ are canonical conjugate form fields, $\Lambda$
is a potential. Let $q$ be a $f$-form and $\epsilon=(-1)^{f}$. The
corresponding Hamiltonian 3-form is defined as follows
\begin{equation}
{\cal{}H}(N):=\pounds_{N}q\wedge{}p-i_{N}{\cal{}L}.\label{28aApril2015}
\end{equation}
Taking the interior product of the Lagrangian density
\begin{eqnarray}
i_{N}{\cal{}L}=\pounds_{N}q\wedge{}p-\epsilon{}i_{N}q\wedge{}dp-\epsilon{}dq\wedge{}i_{N}p
-i_{N}\Lambda-d(i_{N}q\wedge{}p),\label{28bApril2015}
\end{eqnarray}
where the Lie derivative $\pounds_{N}=i_{N}d+di_{N}$. Using
(\ref{28bApril2015}), rewrite (\ref{28aApril2015})
\begin{equation}
{\cal{}H}(N)=N^{\mu}{\cal{}H}_{\mu}+d{\cal{}B}(N),
\end{equation}
where $N^{\mu}{\cal{}H}_{\mu}=\epsilon{}i_{N}q\wedge{}dp
+\epsilon{}dq\wedge{}i_{N}p+i_{N}\Lambda$ which is proportional to
the field equations and vanishes `on shell'.  Note that $N^{\mu}$
is the vector field.  The Hamiltonian density ${\cal{H}}_{\mu}$
determines the evolution equations and initial value constraints.
The natural boundary term ${\cal{}B}(N)=i_{N}q\wedge{}p$. However,
this boundary term is not unique since it can be removed by
introducing a new Hamiltonian
\begin{equation}
{\cal{}H'}(N)={\cal{}H}(N)+d(-i_{N}q\wedge{}p)
=N^{\mu}{\cal{}H}_{\mu}.
\end{equation}
Taking the variation of this new Hamiltonian
\begin{equation}
\delta{\cal{}H'}(N)=-i_{N}({\rm{}F.E.})
-\delta{}q\wedge\pounds_{N}p
+\pounds_{N}q\wedge\delta{}p+d{\cal{}B}(N),
\end{equation}
where the field equation
${\rm{}F.E.}=\delta{}q\wedge\frac{\delta{\cal{}L}}{\delta{}q}
+\frac{\delta{\cal{}L}}{\delta{}p}\wedge\delta{}p$. The boundary
variation term is
\begin{equation}\label{16}
{\cal{}B}(N)=-i_{N}q\wedge\delta{}p
+\epsilon\delta{}q\wedge{}i_{N}p.
\end{equation}
This ${\cal{B}}(N)$ cannot be removed because it comes from
$\delta{\cal{}H'}$ directly.  Boundary conditions can be obtained
through the boundary term in the variation of the Hamiltonian
vanishes. We add an appropriate boundary term to the Hamiltonian
\begin{equation}
{\cal{}H'}(N)\rightarrow{\cal{}H}_{k}(N)
=N^{\mu}{\cal{}H}_{\mu}+d{\cal{}B}_{k}(N),
\end{equation}
to modify the variational boundary term. In order to achieve nice
components like $i_{N}(\delta{}q\wedge\Delta{}p)$ or
$i_{N}(\Delta{}q\wedge\delta{}p)$, there are four simple boundary
expressions can be added. The variation of the four Hamiltonians
including this four expressions are
\begin{eqnarray}
\delta{\cal{}H}_{q}(N)=K +di_{N}(\delta{}q\wedge\Delta{}p),&&
\delta{\cal{}H}_{d}(N)=K -d(i_{N}\Delta{}q\wedge\delta{}p-\epsilon\delta{}q\wedge i_{N}\Delta p),\\
\delta{\cal H}_{p}(N)=K-di_{N}(\Delta{}q\wedge\delta{}p),&&
\delta{\cal{}H}_{c}(N)=K+d(i_{N}\delta{}q\wedge\Delta{}p-\epsilon\Delta{}q\wedge{}i_{N}\delta{}p),
\end{eqnarray}
where $K=-i_{N}({\rm{}F.E.})
-\delta{}q\wedge\pounds_{N}p+\pounds_{N}q\wedge\delta{}p$. Thus we
recovered~\cite{Nester1999}
\begin{eqnarray}
{\cal{}B}_{q}(N)=i_{N}q\wedge\Delta{}p-\epsilon\Delta{}q\wedge{}i_{N}\overline{p},
&&{\cal{}B}_{d}(N)=i_{N}\overline{q}\wedge\Delta{}p-\epsilon\Delta{}q\wedge{}i_{N}\overline{p},\\
{\cal{}B}_{p}(N)=i_{N}\overline{q}\wedge\Delta{}p-\epsilon\Delta{}q\wedge{}i_{N}{p},
&&{\cal{}B}_{c}(N)=i_{N}q\wedge\Delta{}p-\epsilon\Delta{}q\wedge{}i_{N}{p},
\end{eqnarray}
where $\Delta{}q=q-\overline{q}$, $\Delta{}p=p-\overline{p}$, both
$\overline{q}$ and $\overline{p}$ are the background reference
values. Alternatively, rewrite the above four equations in a
compact form
\begin{equation}
{\cal B}_{k_{1},k_{2}}(N)={\cal B}_{p}(N)
+k_{1}i_{N}\Delta{}q\wedge\Delta{}p+\epsilon
k_{2}\Delta{}q\wedge{}i_{N}\Delta{}p,
\end{equation}
where $k_{1}$ and $k_{2}$ can be $0$ or $1$. In detail
${\cal{}B}_{0,0}={\cal B}_{p}$,
${\cal{}B}_{0,1}={\cal{}B}_{\rm{}d}$,
${\cal{}B}_{1,0}={\cal{}B}_{\rm{}c}$ and
${\cal{}B}_{1,1}={\cal{}B}_{q}$.

Using the analogy of classical electrodynamics and apply to the
relativistic gravity, the type of boundary condition should be
either Dirichlet or Neumann, and even a mixture of these two. From
the boundary condition point of view, we prefer ${\cal{}B}_{q}$
and ${\cal{}B}_{p}$ since they are the simplest, i.e., Dirichlet
or Neumann. Meanwhile, the boundary conditions of ${\cal{B}}_{c}$
or ${\cal{B}}_{d}$ could be a certain linear combination of
Dirichlet and Neumann. For the case of ${\cal{}B}_{q}$, there are
two ways to obtain the variation of the boundary term that satisfy
$i_{N}(\delta{}q\wedge\Delta{}p)=0$. First control $q$, then
$\delta{}q=0$. The second is to freely vary $q$ and then
$\delta{}q$ becomes arbitrary, which implies $\Delta{}p=0$. This
is the `natural boundary condition' as it forces $p=\overline{p}$.
Similarly for the variation boundary term
$i_{N}(\Delta{}q\wedge\delta{}p)=0$. In addition, we have modified
$(k_{1},k_{2})\rightarrow(c_{1},c_{2})$, where $c_{1}$ and $c_{2}$
are arbitrary constants, such that the variation is still
legitimate~\cite{SoIJMPD}.

\subsection{Quasi-local M$\o$ller and Freud super-potentials}
Here we apply this Hamiltonian formalism to the Einstein-Hilbert
Lagrangian~\cite{Nester1999}
\begin{eqnarray}
{\cal{L}}_{\rm{}GR}:=R^{\alpha}{}_{\beta}\wedge\eta_{\alpha}{}^{\beta},
\end{eqnarray}
where the curvature 2-form
$R^{\alpha}{}_{\beta}=d\Gamma^{\alpha}{}_{\beta}+\Gamma^{\alpha}{}_{\gamma}\wedge\Gamma^{\gamma}{}_{\beta}$,
the connection 1-form
$\Gamma^{\alpha}{}_{\beta}=\Gamma^{\alpha}{}_{\beta\gamma}dx^{\gamma}$
and the dual basis is
$\eta^{\alpha\beta\cdots}=*(dx^{\alpha}\wedge{}dx^{\beta}\cdots)$.
As before, the interior product
\begin{eqnarray}
i_{N}{\cal{L}}=i_{N}R^{\alpha}{}_{\beta}\wedge\eta_{\alpha}{}^{\beta}
+R^{\alpha}{}_{\beta}\wedge{}i_{N}\eta_{\alpha}{}^{\beta}
=\pounds_{N}\Gamma^{\alpha}{}_{\beta}\wedge\eta_{\alpha}{}^{\beta}-{\cal{H}}(N).
\end{eqnarray}
After a straightforward manipulation, the Hamiltonian density from
above becomes
\begin{eqnarray}
{\cal{H}}(N)=N^{\mu}{\cal{H}}_{\mu}-i_{N}\Gamma^{\alpha}{}_{\beta}D\eta_{\alpha}{}^{\beta}
+d{\cal{B}}(N),\label{6aMay2015}
\end{eqnarray}
where ${\cal{H}}_{\alpha}=2G^{\rho}{}_{\alpha}\eta_{\rho}$ which
satisfies the dynamical evolution and initial value constraints,
$i_{N}\Gamma^{\mu}{}_{\nu}\wedge{}D\eta_{\mu}{}^{\nu}$ vanishes
since the metric compatibility. Finally the boundary term
${\cal{B}}=i_{N}\Gamma^{\mu}{}_{\nu}\wedge\eta_{\mu}{}^{\nu}
=N^{\alpha}g^{\nu\sigma}\Gamma^{\mu}{}_{\sigma\alpha}\eta_{\mu\nu}$.
Note that it is legal to modify the boundary term replace a
negative sign, i.e., ${\cal{B}}\rightarrow-{\cal{B}}$.  Rewrite
(\ref{6aMay2015})
\begin{eqnarray}
{\cal{H}}(N)=N^{\alpha}\left[2G^{\rho}{}_{\alpha}\eta_{\rho}
-\frac{1}{2}\sqrt{-g}(g^{\nu\sigma}\Gamma^{\mu}{}_{\sigma\alpha}
-g^{\mu\sigma}\Gamma^{\nu}{}_{\sigma\alpha})dS_{\mu\nu}
\right],\label{6bMay2015}
\end{eqnarray}
where
$dS_{\mu\nu}=\frac{1}{2}\epsilon_{\mu\nu\xi\kappa}dx^{\xi}\wedge{}dx^{\kappa}$.
Looking at (\ref{6bMay2015}), the term
$_{\rm{}M}{\cal{}U}_{\alpha}{}^{[\mu\nu]}=2\sqrt{-g}g^{\sigma[\mu}\Gamma^{\nu]}{}_{\sigma\alpha}$
looks like M$\o$ller super-potential~\cite{Moller} but not
exactly. Instead we call this quasi-local Moller super-potential.
The reason is that the M$\o$ller pseudo-tensor cannot fulfill the
interior mass-energy requirement $2G^{\rho}{}_{\mu}$ as explained
before, but the quasi-local M$\o$ller super-potential
$_{\rm{}M}{\cal{U}}_{\alpha}{}^{[\mu\nu]}$ can. The variation of
(\ref{6aMay2015})
\begin{eqnarray}
\delta{\cal{H}}={\rm{}field~equation~terms}
+di_{N}(\delta\Gamma^{\alpha}{}_{\beta}\wedge\eta_{\alpha}{}^{\beta}).
\end{eqnarray}
This can be classified as Neumann type boundary condition, i.e.,
the connection $\Gamma\simeq\partial{}g$ is to be held fixed.
Alternatively this terminology can be translated as the deDonder
gauge:
$0=\partial_{\kappa}(\sqrt{-g}g^{\xi\kappa})=\sqrt{-g}\Gamma^{\xi\kappa}{}_{\kappa}$.
Using the analogy of the pseudo-tensor method, the quasi-local
M$\o$ller pseudo-tensor is
$_{\rm{}M}\mathbf{t}_{\alpha}{}^{\mu}=\partial_{\nu}({}_{\rm{}M}{\cal{U}}_{\alpha}{}^{[\mu\nu]})$.
It is known that the tidal heating is
$\dot{W}_{\rm{}M}=-\frac{1}{2}\dot{I}_{ij}E^{ij}$, where
$\dot{E}_{\rm{int}}=\frac{\beta+2}{10}\frac{d}{dt}(I_{ij}E^{ij})=0$
which means the energy localization chosen
$\beta=-2$~\cite{Favata}.

Keep the same track in~\cite{Nester1999}, swap the quasi-local
M$\o$ller super-potential to the other pattern
${\cal{B}}'=\Gamma^{\mu}{}_{\nu}\wedge{}i_{N}\eta_{\mu}{}^{\nu}=\frac{1}{2}N^{\alpha}g^{\beta\sigma}
\Gamma^{\lambda}{}_{\beta\gamma}\delta^{\gamma\mu\nu}_{\lambda\sigma\alpha}\eta_{\mu\nu}$.
Again flap the sign of ${\cal{B}}'$ and rewrite (\ref{6aMay2015})
\begin{eqnarray}
{\cal{H}}(N)=N^{\alpha}\left[2G^{\rho}{}_{\alpha}\eta_{\rho}
-\frac{1}{2}\sqrt{-g}g^{\beta\sigma}\Gamma^{\lambda}{}_{\beta\gamma}
\delta^{\gamma\mu\nu}_{\lambda\sigma\alpha} dS_{\mu\nu}
\right],\label{6cMay2015}
\end{eqnarray}
where
$_{\rm{}F}{\cal{}U}_{\alpha}{}^{[\mu\nu]}=-\sqrt{-g}g^{\beta\sigma}\Gamma^{\lambda}{}_{\beta\gamma}
\delta^{\gamma\mu\nu}_{\lambda\sigma\alpha}$ looks like the Freud
super-potential~\cite{Einstein} but we call this the quasi-local
Freud super-potential because it comes from the Hamiltonian
formalism. The quasi-local Einstein pseudotensor is
$_{\rm{}E}\mathbf{t}_{\alpha}{}^{\mu}=\partial_{\nu}(_{\rm{}F}{\cal{U}}_{\alpha}{}^{[\mu\nu]})$.
The known result for the tidal heating
$\dot{W}_{\rm{}E}=\frac{3}{10}\frac{d}{dt}(I_{ij}E^{ij})-\frac{1}{2}\dot{I}_{ij}E^{ij}$,
where
$\dot{E}_{\rm{int}}=\frac{\beta+2}{10}\frac{d}{dt}(I_{ij}E^{ij})=0$
which means the energy localization selected
$\beta=1$~\cite{Favata}. Note that this boundary condition can be
described as the Dirichlet type which means fixing
$\sqrt{-g}g^{\beta\sigma}$.

Based on the pseudo-tensor method, Thorne claimed that the tidal
heating is unique~\cite{Thorne} and Purdue verified that indeed it
is gauge-invariant~\cite{Purdue}. As far as the tidal heating is
concerned, we find that the gauge condition and boundary condition
are equivalent. Moreover we are going to verify that all the
quasi-local boundary expressions obtain the standard tidal heating
rate.

\subsection{Relativistic quasi-local boundary expressions}
Here we write the modified quasi-local expressions in holonomic
frames~\cite{SoIJMPD}
\begin{eqnarray}
{\cal{B}}(N)&=&{\cal{B}}_{p}(N)+c_{1}i_{N}\Delta\Gamma^{\alpha}{}_{\beta}\wedge\Delta\eta_{\alpha}{}^{\beta}
-c_{2}\Delta\Gamma^{\alpha}{}_{\beta}\wedge{}i_{N}\Delta\eta_{\alpha}{}^{\beta}\nonumber\\
&=&-\frac{1}{2}N^{\alpha}\left(_{\rm{}F}{\cal{}U}_{\alpha}{}^{[\mu\nu]}
+c_{1}\sqrt{-g}h^{\lambda\pi}\Gamma^{\sigma}{}_{\alpha\pi}\delta^{\mu\nu}_{\lambda\sigma}
+c_{2}\sqrt{-g}h^{\beta\sigma}\Gamma^{\tau}{}_{\lambda\beta}\delta^{\lambda\mu\nu}_{\tau\sigma\alpha}\right)
\epsilon_{\mu\nu},\label{10aApril2015}
\end{eqnarray}
where
$\Delta\Gamma^{\alpha}{}_{\beta\mu}=\Gamma^{\alpha}{}_{\beta\mu}-\bar{\Gamma}^{\alpha}{}_{\beta\mu}$,
$c_{1},c_{2}$ are real and finite. For simplicity, consider the
reference for flat spacetime
$\bar{\Gamma}^{\alpha}{}_{\beta\mu}=0$ in Cartesian coordinates.
Extract the quasi-local superpotential in (\ref{10aApril2015})
\begin{eqnarray}
{\cal{}U}_{\alpha}{}^{[\mu\nu]}={}_{\rm{}F}{\cal{}U}_{\alpha}{}^{[\mu\nu]}
+c_{1}\sqrt{-g}h^{\lambda\pi}\Gamma^{\sigma}{}_{\alpha\pi}\delta^{\mu\nu}_{\lambda\sigma}
+c_{2}\sqrt{-g}h^{\beta\sigma}\Gamma^{\tau}{}_{\lambda\beta}\delta^{\lambda\mu\nu}_{\tau\sigma\alpha}
\end{eqnarray}
Bear in mind that the Hamiltonian has already fulfilled the inside
matter value $2G_{\alpha}{}^{\beta}$. This quasi-local Freud
super-potential and the extra higher order terms $h\Gamma$ only
contribute the energy-momentum in vacuum. Carry on the calculation
and we have the quasi-local pseudo-tensor
\begin{eqnarray}
\mathbf{t}_{\alpha}{}^{\mu}&=&{}_{\rm{}E}\mathbf{t}_{\alpha}{}^{\mu}
+c_{1}[(\Gamma^{\lambda\pi}{}_{\nu}+\Gamma^{\pi\lambda}{}_{\nu})\Gamma^{\sigma}{}_{\alpha\pi}
+h^{\lambda\pi}\Gamma^{\sigma}{}_{\alpha\pi,\nu}]\delta^{\mu\nu}_{\lambda\sigma}\nonumber\\
&&+c_{2}[(\Gamma^{\beta\sigma}{}_{\nu}+\Gamma^{\sigma\beta}{}_{\nu})\Gamma^{\tau}{}_{\lambda\beta}
+h^{\beta\sigma}\Gamma^{\tau}{}_{\lambda\beta,\nu}]\delta^{\lambda\mu\nu}_{\tau\sigma\alpha}\nonumber\\
&=&\delta^{\mu}_{\alpha}(\Gamma^{\beta\nu}{}_{\lambda}\Gamma^{\lambda}{}_{\beta\nu}
-\Gamma^{\pi}{}_{\pi\nu}\Gamma^{\nu\beta}{}_{\beta})
+\Gamma^{\beta}{}_{\beta\alpha}(\Gamma^{\mu\nu}{}_{\nu}-\Gamma^{\nu\mu}{}_{\nu})
+(\Gamma^{\beta\mu}{}_{\alpha}+\Gamma^{\mu\beta}{}_{\alpha})\Gamma^{\nu}{}_{\nu\beta}
-2\Gamma^{\beta\nu}{}_{\alpha}\Gamma^{\mu}{}_{\beta\nu}
\nonumber\\
&&+c_{1}\left[\Gamma^{\lambda\beta}{}_{\alpha}\Gamma^{\mu}{}_{\lambda\beta}
+\Gamma^{\beta\nu}{}_{\alpha}\Gamma_{\nu\beta}{}^{\mu}
-\Gamma^{\mu}{}_{\alpha\beta}\Gamma^{\nu\beta}{}_{\nu}
-\Gamma^{\mu}{}_{\alpha\pi}\Gamma^{\pi\nu}{}_{\nu}
+h^{\mu\pi}\Gamma^{\nu}{}_{\alpha\pi,\nu}
-h^{\pi\nu}\Gamma^{\mu}{}_{\alpha\pi,\nu}\right]\nonumber\\
&&+c_{2}\left[
\begin{array}{ccccc}
\delta^{\mu}_{\alpha}(2\Gamma^{\beta\nu}{}_{\lambda}\Gamma^{\lambda}{}_{\beta\nu}
-\Gamma^{\beta}{}_{\beta\lambda}\Gamma^{\lambda\nu}{}_{\nu}
-\Gamma^{\beta}{}_{\beta\lambda}\Gamma^{\nu\lambda}{}_{\nu})
+(2\Gamma^{\mu\beta}{}_{\alpha}+\Gamma^{\beta\mu}{}_{\alpha})\Gamma^{\nu}{}_{\nu\beta}\\
-\Gamma^{\beta\nu}{}_{\alpha}(3\Gamma^{\mu}{}_{\beta\nu}+\Gamma_{\nu\beta}{}^{\mu})
+\Gamma^{\mu}{}_{\alpha\beta}\Gamma^{\beta\nu}{}_{\nu}
-h^{\beta\nu}R^{\mu}{}_{\beta\alpha\nu}\quad\quad\quad\quad\quad\quad~
\end{array}
\right].
\end{eqnarray}
Consequently the tidal heating is
\begin{eqnarray}
\dot{W}=\frac{3+2c_{2}}{10}\frac{d}{dt}(I_{ij}E^{ij})-\frac{1}{2}\dot{I}_{ij}E^{ij}.
\end{eqnarray}
This shows that the term with $c_{2}$ contribute non-vanishing
$\dot{E}_{\rm{}int}$.  In contrast, the term with $c_{1}$
contributes nothing. Hence, the tidal heating rate is indeed
boundary conditions independent.

\section{Conclusion}
Purdue and Favata calculate the tidal heating used classical
pseudo-tensors.  Booth and Creighton employed the quasi-local mass
formalism of Brown and York to demonstrate the same subject. All
of them give the result matched with the Newtonian gravity.  Here
we present another Hamiltonian quasi-local boundary expressions
and all give the same desired result. This illustrates that the
tidal heating is unambiguous since it is gauge invariant. In fact
the ambiguity comes from $E_{\rm{}int}$: energy interaction
between the isolated body quadrupolar deformation and
 external tidal field.

Meanwhile one may argue that this tidal heating uniqueness is
natural because the principle of the pseudo-tensor method and
quasi-local method are essentially equivalent. Although the main
purpose of these two methods is to evaluate the tidal heating and
obtain the same value, we discovered that they are different
fundamentally. We prefer the Hamiltonian quasi-local formalism
since it guarantees the dynamical evolution and initial value
constraints, while the pseudo-tensor method cannot warranty the
inside matter requirement.


\end{document}